\documentclass[2col]{aa}

\usepackage{graphicx}
\usepackage{amsmath}
\usepackage{wasysym}
\usepackage{breakcites}
\usepackage{comment}
\usepackage[symbol]{footmisc}
\usepackage{txfonts}

\usepackage{lineno}

\title{Convective inhibition with an ocean}
\subtitle{I: supercritical cores on sub-Neptunes/super-Earths}

\author{S. Markham
\inst{1}
\and
T. Guillot
\inst{1}
\and
D. Stevenson
\inst{2}}

\institute{Universit\'{e} C\^{o}te d'Azur, Observatoire de la C\^{o}te d'Azur, CNRS, Laboratoire Lagrange, France\\
\email{steve.markham@oca.eu}
\and
California Institute of Technology, Dept. of Geological and Planetary Sciences, Pasadena, CA\\
}
\date{\today}

\begin{document}

\abstract
{}
{In this work we generalize the notion of convective inhibition to apply it to cases where there is an infinite reservoir of condensible species (i.e., an ocean). 
We propose a new model for the internal structure and thermal evolution of super-Earths with hydrogen envelopes.}
{We derive the criterion for convective inhibition in a generalized phase mixture from first principles thermodynamics.
We then investigate the global ocean case using a water-hydrogen system, for which we have data, as an example. 
After illustrating the relevant thermodynamics, we extend our arguments to  apply to a system of hydrogen and silicate vapor. 
We then employ a simple atmospheric model to apply our findings to super-Earths and to make predictions about their internal structures and thermal evolution.}
{For hydrogen envelope masses roughly in the range  $10^{-3}-10^{-1}M_\oplus$, convective contact between the envelope and core may shut down because of the compositional gradient that arises from silicate partial vaporization. 
For envelope hydrogen masses that cause the associated basal pressure to exceed the critical pressure of pure silicate (on the order of a couple kilobars), the base of that envelope and the top of the core lie on the critical line of the two-phase hydrogen-silicate phase diagram. 
The corresponding temperature is much higher than convective models would suggest. 
The core is then ``supercritical'' in the sense that the temperature exceeds the critical temperature for pure silicate. 
The core then cools inefficiently, with intrinsic heat fluxes potentially comparable to the Earth's internal heat flux today. }
{This low heat flux may allow the core to remain in a high entropy supercritical state for billions of years, but the details of this depend on the nature of the two-component phase diagram at high pressure, something that is currently unknown. 
A supercritical core thermodynamically permits the dissolution of large quantities of hydrogen into the core. 
}

\maketitle

\section{Introduction}
When the molecular weight of a condensing volatile species exceeds the mean molecular weight of the gaseous atmosphere it inhabits, convective inhibition may occur, permitting stably stratified superadiabatic temperature gradients \citep{guillot1995}. 
Applications of convective inhibition to our Solar System has garnered renewed interest in recent years; it has been invoked to explain exotic meteorology on Saturn \citep{li-ingersoll2015}, the underlying fluid dynamical theory has been further scrutinized \citep{friedson-gonzales2017, leconte+2017}, and it has been applied to refine internal states and thermal evolution of Uranus and Neptune \citep{markham-stevenson2021}. 
These models in general take as an input parameter the bulk mixing ratio of the condensing species at depth, assumed to be well mixed and dominated by hydrogen. 
The fundamental physics of the case where the mixture at depth is dominated instead by the condensing species, namely, a low-mass envelope overlying a global ocean, has not yet been considered. 
In this work we will approach the problem using thermodynamic arguments regarding the coexistence of multicomponent systems, generalizing the notion of convective inhibition to apply it to any phase mixture. 
The new insights in this paper are important because the case we consider, a low-mass envelope overlying a global ocean, is expected to correspond to super-Earth and sub-Neptune exoplanets. \\

Sub-Neptune and super-Earths were among the very first observed exoplanets, with Poltergheist and Phobetor found orbiting the pulsar Lich in 1992 \citep{wf1992}. 
The structure and composition of these bodies, intermediate in mass between Earth and Neptune with no analogs in the Solar System, have only gradually come into focus since. 
In the literature these bodies are currently referred to as either sub-Neptunes or super-Earths depending on the context; in this work we preferentially refer to them as ``super-Earths'' for clarity, although we are referring more generally to bodies in the mass range roughly between two and ten Earth masses that retain a gaseous envelope. 
The  deluge of data from the Kepler mission represents the greatest advancement in our understanding of these bodies, making clear that planets do not have to become much larger than Earth before they begin to rapidly increase in apparent size \citep{borucki+2011, furlan+2017} when defining size using the millibar pressure. 
These bodies' mean densities can be described by a composition dominated by water \citep{seager+2007}, although the dearth of observed high-mass, high-density planets suggests many are likely to be a mixture of primarily silicates and hydrogen. 
A small amount of hydrogen can make an enormous difference in a planet's apparent radius, because the millibar level can be many scale heights above a physical surface, and the combination of low molecular mass and high temperatures commonly observed in these bodies imply large scale heights. 
Canonical models usually involve an extended, convective gas envelope in thermal contact with a magma ocean \citep[e.g.,][]{ginzburg+2016, vazan+2018}. 
This model, which involves a silicate core with an overlying gas envelope, will be the focus of this paper. 
In particular, we discuss the implications of convective inhibition as a result of the condensation of silicate vapor. 
This paper is the first in a planned series of papers intended to fundamentally rethink our conception of these bodies, which perhaps behave much more like gas planets than terrestrial analogs with a gas envelope. \\

Recent observations of Jupiter and Saturn indicate both planets possess an extended diffuse core \citep{stevenson2020, mankovich-fuller2021, dongdong2019, miguel+2022}, suggesting that significant metal enrichment may go alongside gas accretion during planetary formation \citep{ormel+2021}. 
In this paper we argue that if the initial envelope of super-Earths is substantially enriched in silicates, as recent formation models contend \citep{bodenheimer+2018, brouwers-ormel2020}, this can meaningfully complicate our understanding of super-Earth structure and evolution. 
We argue that such enrichment leads naturally to a layer of static stability at depth such that the envelope convectively decouples from the core. 
In this work we demonstrate that convective inhibition of this kind is likely to occur on silicate planets that possess an envelope on the order on the order of 0.1\% - 10\% $M_\oplus$.  
The phenomenon we describe may also apply to greater envelope masses, depending on currently unknown details regarding the critical mixtures of hydrogen and silicates. 
Planets with hydrogen envelopes in the mass range 0.1\% - 10\% $M_\oplus$ are probably the most common in the universe \citep{borucki+2011, furlan+2017}. \\

A layer of static stability at depth reduces super-Earth luminosity at early times such that the core loses heat inefficiently. 
Because of the expected low thermal transport efficiency under the relevant conditions (see Sect.~\ref{stable-layer}), we expect the layer of static stability to be thin compared to the full extent of the atmosphere, even though the temperature change across the stable layer can be thousands of kelvins. 
In the thin stable layer limit,  we can model a super-Earth's steady state internal heat flow based on the mass of its envelope. 
We argue that the top of the core of these planets lies on the critical line of the hydrogen--silicate binary system at a temperature and pressure that exceed the corresponding values for the critical point of pure silicates, which we call the ``critical endpoint'' in this work.   
We expect the critical endpoint pressure level to be on the order of a couple kilobars \citep{xiao-stixrude}, corresponding to an envelope mass of roughly 0.1\% $M_\oplus$. 
The corresponding base of the convective part of the envelope can be significantly cooler than the top of the core, such that the planet's internal heat flux can be small even when the core is very hot ($\sim 10^4$K). 
We predict that the luminosity of planets drops rapidly after isolation, but then remains nearly constant over geologically long timescales. 
This is an exploratory work focused primarily on describing a new model for super-Earth internal structure and thermal evolution, and constraining the relevant orders of magnitude. 
We intend to follow up this work with more detailed studies that address questions that naturally arise from this new model, including hydrogen pollution of the core and the process of formation. \\

\begin{figure}[h!]
\vspace{10pt}
\centering
\includegraphics[width=\linewidth]{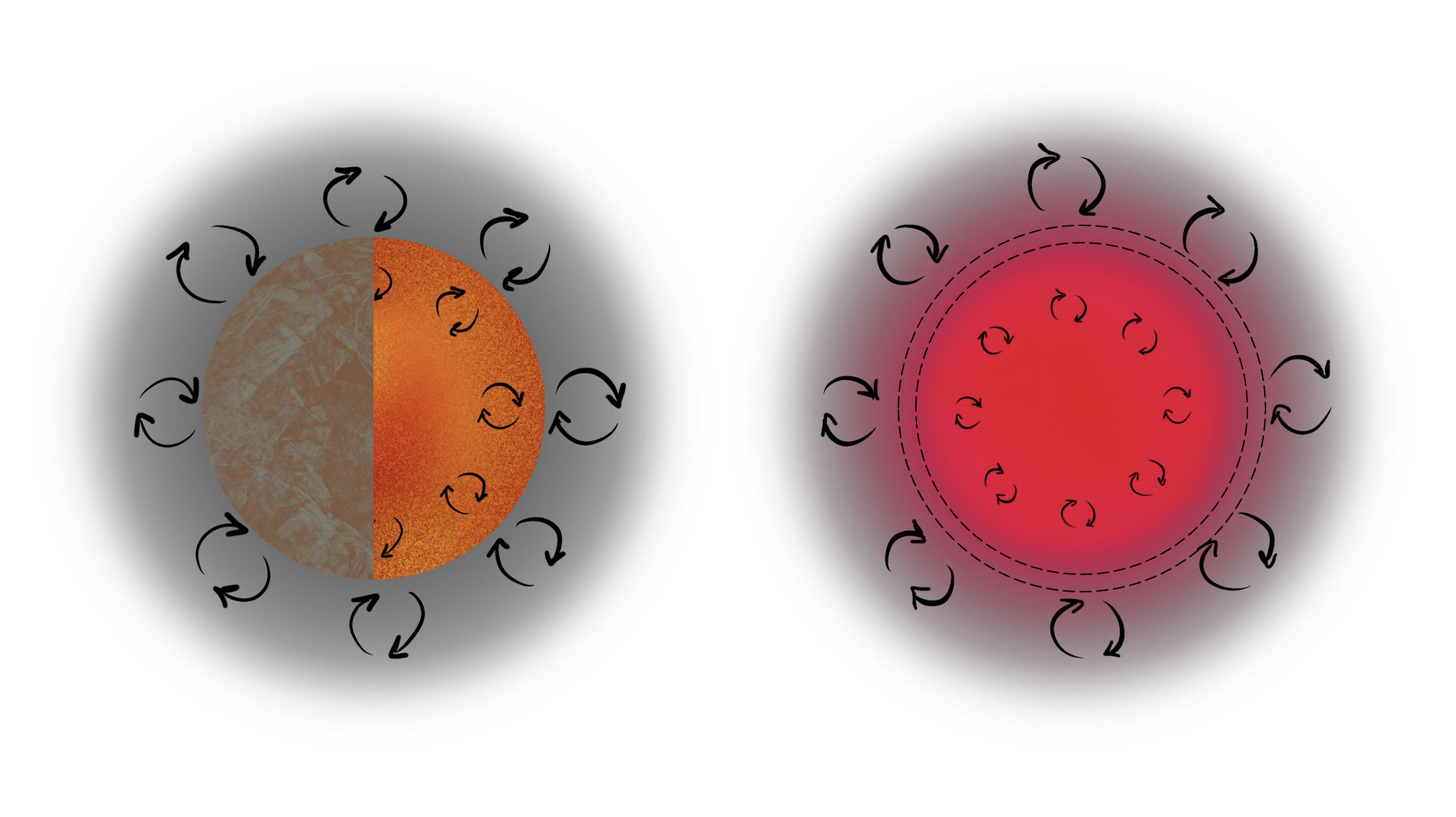}
\caption{Cartoon representation of this paper's findings.  On the left is a representation of the canonical view of super-Earths \citep[e.g.,][]{ginzburg+2016}, with a frozen or molten silicate core in good thermal contact with a convective gas envelope.  On the right we show our proposed model, with a layer of static stability between a (possibly hydrogen-polluted) supercritical fluid core and the silicate-vapor-enriched gas envelope.}
\vspace{10pt}
\label{fig:cartoon}
\end{figure}

This work follows the following structure: 
in Sect.~\ref{sec:theory}, we extend the theory of convective inhibition to generalized thermodynamic contexts, using a water--hydrogen system (for which, unlike silicates, we have data) as an example to illustrate the relevant thermodynamics at play. 
In Sect.~\ref{sec:structure}, we outline the internal structure of the super-Earths we model in this work. 
In Sect.~\ref{sec:evol}, we calculate orders of magnitude for core cooling timescales and argue that under certain conditions the core can retain much of its primordial heat for billions of years. 
Then in Sect.~\ref{sec:followup} we identify several natural consequences of the super-Earth model we present in this work, but that lie beyond the scope of this work, to motivate collection of further data and follow-up papers in this series. 
Finally in Sect.~\ref{sec:discussion} we summarize our findings, comment on observable consequences of our model, and identify the key data necessary to further advance our understanding of the most common class of planets in our galaxy. 


\section{Theoretical foundations}
\label{sec:theory}
In this section we aim to introduce the general notion of convective inhibition for unfamiliar readers, and to generalize previous theoretical work to apply in more exotic contexts. 
This paper follows the preestablished logic of convective inhibition by condensation in hydrogen atmospheres \citep{guillot1995}. 
Prior works focus on planetary atmospheres in the Solar System and assume that the liquid phase remains pure, and that the saturation vapor pressure follows Clausius-Clapeyron with constant latent heat. 
These approximations are appropriate in the context of low pressures and where the mixing ratio of the vapor is much less than unity. 
While these approximations are valid in the context of Solar System atmospheres, they break down at higher pressures and temperatures likely to be relevant on super-Earths. 
Therefore we wish to extend the logic to a general mixture of coexisting phases that remains applicable much deeper into the interior of planets. 
Of particular interest to this work is the case of a relatively low-mass hydrogen atmosphere with an infinite reservoir of volatile condensates (i.e., an ocean). 
This system will be of particular interested because observations indicate a large population of planets with masses intermediate between that of Earth and Neptune with radii far larger than can be explained with an iron--silicate composition \citep[e.g.,][]{furlan+2017}. 
If we adopt the prevailing interpretation, that the observed radii are due to hydrogen envelopes comprising about 1\% of the planet's mass, then we expect envelope basal pressures to exceed the critical endpoint pressure of silicates (1-2 kbar; \cite{xiao-stixrude}, $M_{\rm atm} \sim 10^{-3} M_\oplus$). 
Under these conditions, it is no longer appropriate to approximate the coexistence of the condensate with the dry gas according to the saturation vapor pressure of the pure condensate (which we refer to as the ideal gas approximation in this work); one must use coexistence data. 
Because this scenario violates the assumptions underpinning prior derivations, a new more generalized framework is needed. \\

In this section we inspect three crucial differences in the regime of interest compared to prior works. 
In Sect.~\ref{sec:proof} we re-derive the phenomenon of convective inhibition in a framework that is robust to generalized thermodynamic environments. 
In Sect.~\ref{sec:latent} we seek to specify the top and bottom boundaries of a stable layer resulting from a critical mixture of a condensing species with gas. 
Then in Sect.~\ref{sec:coexistence} we motivate the necessity for Sect.~\ref{sec:proof} by indicating the importance of empirical coexistence curves for mixtures at supercritical pressures. 
In Sect.s~\ref{sec:latent}~and~\ref{sec:coexistence} we use data for a water/hydrogen system to illustrate the behavior with data, and note that the qualitative behavior for silicates should be similar but scaled to an order of magnitude higher temperatures and pressures. 
\\

\subsection{Convective inhibition in a generalized mixture}
\label{sec:proof}

Convective inhibition for an ideal gas coexisting with a condensing species has been derived several times in the past \citep{guillot+1995, leconte+2017, friedson-gonzales2017}. 
The criterion is often written in terms of mass ratio
\begin{equation}
q_{\rm inh} = \frac{R T}{(\mu_B - \mu_A) L},
\end{equation}
where $\mu_A$ and $\mu_B$ are the molar masses of the dry gas and condensing vapor respectively, and $L$ is the latent heat of vaporization. 
This can be equivalently written in terms of the inhibition molar concentration \citep{markham-stevenson2021}. 
\begin{equation}
\label{eq:criterion_simple}
x_{\text{inh}}^{-1} = \left(\frac{\mu_B L}{R T} - 1 \right)(\epsilon - 1), 
\end{equation}
where $\epsilon \equiv \mu_B / \mu_A$. 
In this section we re-derive this criterion for a generalized phase mixture, making no explicit assumptions about the equation of state of the mixture. 
In the context of this work, we do this to apply the concept of convective inhibition to a critical (non-ideal) mixture; more generally, it can apply to any coexisting species of different densities. 
We put aside the question of the background temperature gradient for now, instead focusing purely on how a temperature increment affects density of an isobaric parcel in thermodynamic equilibrium. 
As we will demonstrate, this approach is sufficient to delineate convective inhibition, and recovers the same result as prior works in the ideal gas limit. \\

The coefficient of thermal expansion is
\begin{equation}
\label{eq:expansion}
\alpha \equiv \frac{1}{V} \left(\frac{\partial V}{\partial T}\right)_{p,x},
\end{equation}
where $V$ is the molar specific volume of the material, and $x$ refers to the composition. 
We postulate that convection operates as normal (heat rises) when the coefficient of thermal expansion $\alpha$ is positive. 
On the other hand, materials with a negative coefficient of thermal expansion will be stable against convection even for a superadiabatic temperature gradient. 
That is the reason why, for example, water in a pond will freeze from the top --- the coefficient of thermal expansion for water is negative between its freezing point and about $4^{\circ}C$, where it is near zero. 
To assess when convection can be shut off in planetary interiors, we seek the point at which the coefficient of thermal expansion becomes zero. 
However, for most pure substances in the universe, the coefficient of thermal expansion is always positive; liquid water is a special case. 
For a pure ideal gas, for example, the coefficient of thermal expansion $\alpha = 1/T > 0 ~~ \forall~ T$. 
As has been demonstrated before, however, convection can be shut off for an ideal gas if its composition can be affected by temperature. 
Therefore we define the ``effective'' coefficient of thermal expansion as
\begin{equation}
\label{eq:effective}
\alpha_{\rm{eff}} \equiv \frac{\mu}{V} \frac{d (V/\mu)}{dT}
\end{equation} 
which accounts for the possibility that the composition and therefore the mean molecular weight $\mu$ of a fluid parcel may also depend on temperature. 
This applies to, for example, a gas parcel in contact with a volatile liquid ocean in thermodynamic equilibrium.  
In that case, increasing its temperature allows additional vapor molecules to enter the parcel, while decreasing its temperature will cause vapor molecules to condense and eventually rain out. 
It can also apply more generally to a variety of contexts relevant to planetary interiors, for example a eutectic system with a dense phase that freezes out, two partially miscible liquids of different densities, or an equilibrium chemical reaction. 
This work will specifically focus on the case of a critical mixture (see Sect.~\ref{sec:coexistence}), although the derivation in this section is more general. 
\\

We now consider an isobaric mixture of two species $A$ and $B$ coexisting in a parcel of fluid, where the molecular mass of species $B$ is greater than that of species $A$, $\mu_B > \mu_A$. 
The parcel contains a fixed mass of species $A$ that does not undergo any phase transitions, in contact with an infinite reservoir of species $B$. 
Species $B$ can enter or leave the system according to its equilibrium concentration. 
\\

In our system the molar concentration 
$x \equiv N_B/(N_A + N_B)$
uniquely determines the composition of the parcel, and its mean molecular weight $\mu = \mu_A(1-x) + \mu_B x$. 
Since $N_A$ is fixed, $x = 0$ when $N_B = 0$ (pure substance $A$), and $x \rightarrow 1$ as $N_B \rightarrow \infty$ (when $B$ becomes completely soluble). 
It follows directly from Eq.~\ref{eq:effective} that the effective coefficient of thermal expansion for this system is 
\begin{equation}
\label{eq:effective2}
\alpha_{\rm{eff}} = \frac{1}{V} \left(\frac{\partial V}{\partial T} \right)_x + \frac{1}{V}\frac{dx}{dT} \left(\frac{\partial V}{\partial x} \right)_T+\frac{dx}{dT}\frac{1-\epsilon}{1+x(\epsilon-1)}.
\end{equation}
It should be noted that the first term on the right-hand side of Eq.~\ref{eq:effective2} is equivalent to the classical coefficient of thermal expansion (Eq.~\ref{eq:expansion}), while the remaining terms are corrections to account for changes in composition. 
The second term on the right-hand side accounts for the possible change in mean molecular spacing that occurs as a result of changing the composition, possibly due to new intermolecular forces between species $A$ and $B$. 
The third and final term on the right-hand side accounts for the change in mean molecular weight associated with changing composition; this term can be negative if $\epsilon > 1 \impliedby \mu_B > \mu_A$. \\

Convection is inhibited if $\alpha_{\rm{eff}} \rightarrow 0$. 
Then rewriting Eq.~\ref{eq:effective2}, we obtain the generalized criterion for convective inhibition. 
\begin{equation}
\label{criterion}
\frac{dx}{dT} = -\alpha(T) \left[\left(\frac{\partial \ln V}{\partial x} \right)_T + \frac{1 - \epsilon}{1+x(T)(\epsilon-1)} \right]^{-1}.
\end{equation}
If one specifies a differentiable miscibility $x(T)$, a deviation from volume additivity $\left(\frac{\partial \ln V}{\partial x} \right)_T$, and a classical coefficient of thermal expansion $\alpha$, one can solve Eq.~\ref{criterion} for $T$, the temperature at which convection will be shut off in equilibrium. 
Substituting this temperature into $x(T)$ implies $x_{\rm inh}$, the inhibition concentration,\footnote{Prior works \citep[e.g.,][]{guillot+1995, friedson-gonzales2017, leconte+2017, markham-stevenson2021} referred to this quantity as the ``critical'' mixing ratio but here we call it the ``inhibition'' concentration so as not to confuse it with the critical curve demarcating where the fluid becomes a supercritical state of matter, which is also of central importance in the regime of interest. } 
if a solution exists. 
If no solution within the domain $0 < x < 1$ exists, then convective inhibition does not occur. 
As an example, we can solve for $x_{\rm inh}$, for an ideal gas $A$ in contact with a partially vaporized volatile liquid $B$ we can specify $\alpha = 1/T$, $\left(\frac{\partial \ln V}{\partial x} \right)_T = 0$, and $\frac{d \ln x}{d \ln T} = \frac{L \mu_B}{RT}$ (Clausius-Clapeyron), where $L$ is the latent heat of vaporization for species $B$ and $R$ is the ideal gas constant. 
In that case we recover Eq.~\ref{eq:criterion_simple}, 
identical to previous works. 
\\

The exact value of the inhibition concentration apparently depends on temperature, but we note for the convenience of the reader that it is on the order of  0.1--1\% in the thermodynamic regimes of interest. 
For all saturated concentrations above the inhibition mixing ratio, superadiabatic temperature gradients will be Ledoux stable against convection. 
\\

We have so far not fully justified why it is sufficient to consider only the isobaric coefficient of thermal expansion. 
To do this we now perform simple linear stability analysis of a parcel in an environment that is in saturated thermodynamic equilibrium with an imposed temperature gradient $\nabla_T = \frac{d \ln T}{d \ln p}$. 
We want to assess whether a parcel displaced upward will be more or less dense than its surroundings. 
For the environment, 
\begin{equation}
\label{eq:environment}
\frac{d \ln (1/\rho_{\rm env})}{d \ln p} = \left( \frac{\partial \ln (1/\rho)}{\partial \ln p} \right)_T + \alpha_{\rm eff} \nabla_T
\end{equation}
For the parcel,
\begin{equation}
\label{eq:parcel}
\frac{d \ln (1/\rho_{\rm parcel})}{d \ln p} = \left( \frac{\partial \ln (1/\rho)}{\partial \ln p} \right)_T + \alpha_{\rm eff} \nabla_{\rm ab}
\end{equation}
where $\nabla_{ab}$ is the adiabatic temperature gradient that can be specified and may take different forms depending on the Gr\"{u}neisen parameter of the materials, latent heat released, etc. 
A convective instability occurs if after some small upward displacement $- \delta p$, $\rho_{\rm parcel} < \rho_{\rm env}$. 
Therefore by subtracting Eq.~\ref{eq:parcel} from Eq.~\ref{eq:environment}, the condition for a convective instability is 
\begin{equation}
\alpha_{\rm eff} (\nabla_T - \nabla_{\rm ab}) > 0
\end{equation}
(cf. Eq. 2 of \cite{guillot+1995} or Eq. 15 of \cite{leconte+2017}). 
Therefore, to tell whether a superadiabatic temperature gradient $\nabla_T > \nabla_{\rm ab}$ will be stable against convection in an inviscid medium, it is sufficient to only calculate the sign of the isobaric effective coefficient of thermal expansion $\alpha_{\rm eff}$ (Eq.~\ref{eq:effective2}); the results are independent of the precise form of $\nabla_T$ and $\nabla_{\rm ab}$ and also independent of the material's compressibility. 
\\

One may question whether the kinetics and dynamics of these processes are relevant, and whether one can truly make the assumption that the system is in thermodynamic equilibrium.  
These are reasonable objections that have been investigated in detail in prior works \citep{friedson-gonzales2017, leconte+2017}. 
These works find convective inhibition to be robust with finite settling timescales and imperfect molecular transport, although there are limits. 
Detailed microphysical modeling is beyond the scope of this work, and we encourage interested readers to refer to the above cited works for a more detailed analysis of these effects.

\subsection{Boundaries of the stable layer}
\label{sec:latent}
We seek the top and bottom boundaries of the stable layer. 
Evidently the inhibition mixing ratio at the top boundary depends only on the local gradient of the condensate--gas coexistence curve. 
Under the ideal gas approximation at constant pressure, then, $x_{\rm inh}$ becomes a function of temperature only. 
At higher pressures, an empirical coexistence curve must be used. 
Now that we have identified the mixing ratio at which convection is shut off, we wish to determine at which mixing ratio the convection reasserts itself.  
For example, in \cite{markham-stevenson2021} this occurs after the mixing ratio reaches its ``bulk'' value, called $q_{\text{max}}$. 
However, this bulk value as such only applies in the limit where the system is dominated by gases. 
In the case we consider now with an infinite reservoir of condensing species, we must determine the mixing ratio at depth by other means. 
The solution depends on the thermodynamic conditions. \\

In this section, we refer to ``subcritical'' and ``supercritical'' conditions by referring to the critical endpoint of the pure condensate. 
We use the notation $(p_{\rm crit}^*, T_{\rm crit}^*)$ to refer to the critical endpoint of the pure condensing substance, to distinguish it from the critical curve $(p_{\rm crit}, T_{\rm crit})(x)$ which depends on the mixing ratio of condensate to hydrogen. 
The endpoint of the critical curve where $x=1$ corresponds to the aforementioned critical endpoint. \\

We first consider the case where the hydrostatic pressure is subcritical. 
In this case, there exists a temperature for which the saturation vapor pressure approaches the hydrostatic pressure. 
Physically, this situation manifests as a stable layer truncated from below by a liquid ocean. \\

\begin{figure}[h!]
\vspace{10pt}
\centering
\includegraphics[width=1.1\linewidth]{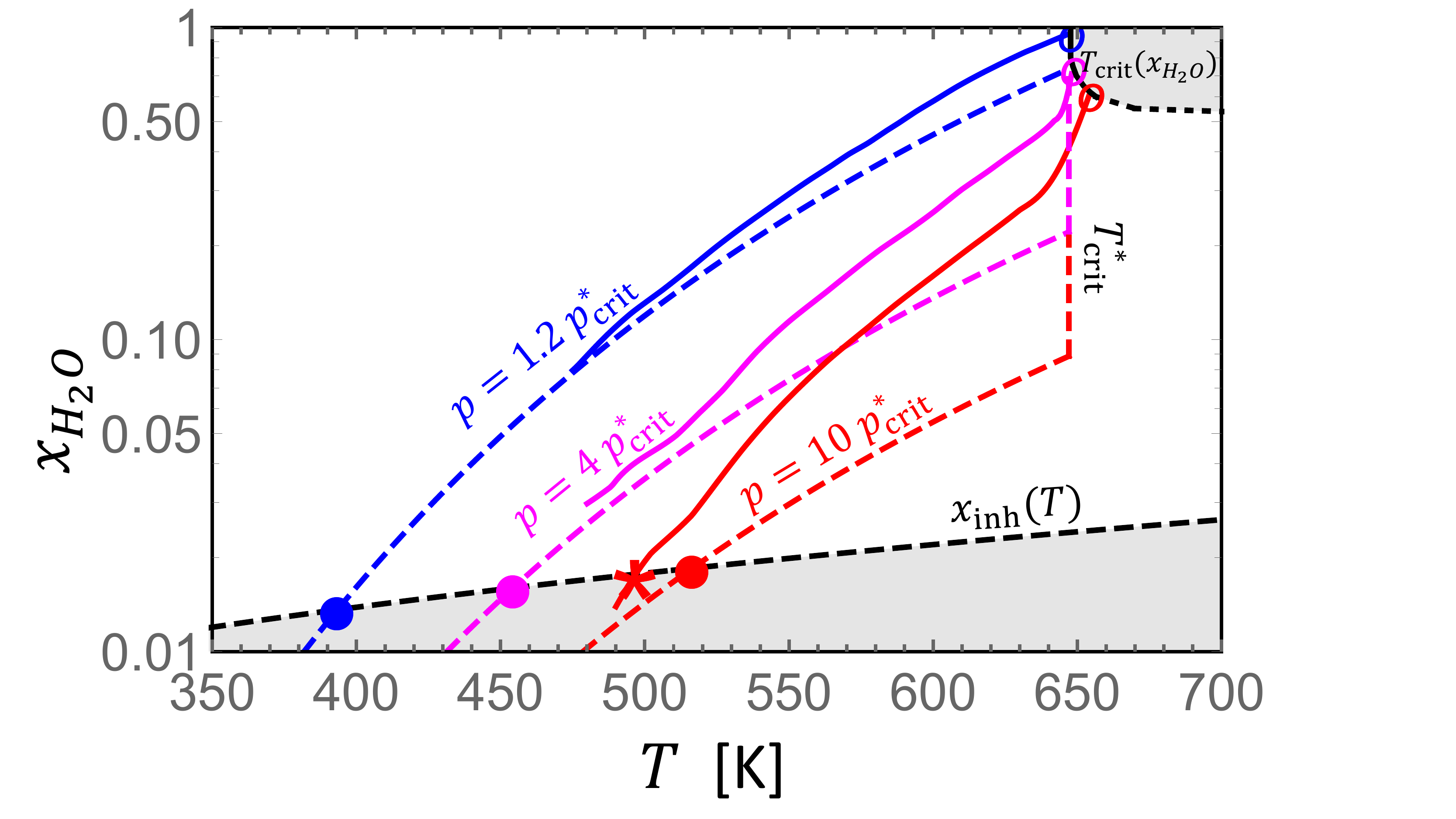}
\caption{Illustration showing the stable layer boundaries in $(x, T)$ space. 
Solid curves use data \citep{seward-franck1989}, see Sect.~\ref{sec:coexistence}), while dashed curves use the ideal gas approximation $x=p_s/p$, diverging near the critical endpoint. 
Gray shaded areas are regions of parameter space wherein free convection is permitted. 
In the top-right corner, the gray region lies beyond the critical curve where phase separation does not occur. 
At the bottom of the figure, the gray region lies below the $x_{\text{inh}}$ curve described in Eq.~\ref{eq:criterion_simple}. 
Blue, magenta, and red curves represent hydrostatic pressure environments of 300 bars, 1 kbar and 2.5 kbars respectively. 
The solid points where the dashed curves intersect correspond to the top of the stable layer under the ideal gas approximation; the red asterisk shows the offset intersection using data. 
The open circles where the solid colored curves intersect the critical curve represent the top of the core. }
\vspace{10pt}
\label{fig:intersection}
\end{figure}

If the hydrostatic pressure is supercritical, then the behavior is subtler. 
In this case, the condensate abundance will increase with increasing temperature, stabilizing against convection, until exceeding the critical temperature, at which point the condensate and gas can mix in all proportions. 
When the two substances can mix in all proportions, condensation no longer occurs so that convective inhibition is no longer at play, and free convection can resume. 
Fig.~\ref{fig:intersection} shows the limits of the ideal gas approximation when dealing with supercritical conditions. 
\\

We see from Fig.~\ref{fig:intersection} the behavior of a coexisting system in pressure environments exceeding the critical endpoint pressure $p_{\rm crit}^*$. 
For marginally supercritical hydrostatic pressures, the ideal gas approximation is appropriate for determining the top of the stable layer, and the bottom of the stable layer is near the critical endpoint. 
However, we see for sufficiently high pressures (red curve in Fig.~\ref{fig:intersection}) that deviations away from the ideal case become important, suggesting somewhat lower temperatures at the top of the stable layer and higher temperatures at the top of the core. 
Hydrogen enrichment at the top of the core may then be significant. 
Indeed, hydrogen enrichment at the top of the core may be quite substantial, approaching a 1:1 molecular mixture at high pressures. 
Following figures will include the same symbols for the important intersection points. 

\subsection{Critical coexistence of rock and gas}
\label{sec:coexistence}
The approximation that $x_s(T) = p_s(T) / p$ where $p_s$ is the saturation vapor pressure and follows an Arrhenius relationship in temperature with constant latent heat is a fair approximation when $p_s \ll p \ll p_{\rm crit}^*$. 
However, in cases where the saturation vapor pressure exceeds the hydrostatic pressure or approaches the critical pressure, as is of interest in this work, such an approximation is no longer valid, and empirical data are necessary to model the coexistence of phases as shown in Fig.~\ref{fig:intersection}. 
The solid curves in Fig.~\ref{fig:intersection} show empirical coexistence curves of a gas--liquid mixture of water and hydrogen \citep{seward-franck1989, bailey-stevenson2021}. 
At low pressures, the liquid phase remains nearly pure, while more and more water enters the gas phase as temperature increases. 
Above the critical temperature of water $\sim$650K, the phases no longer separate and both species can coexist in a well-mixed supercritical fluid. 
Fig.~\ref{fig:intersection} shows the deviation of this empirical data compared to the assumption that $p_s = p_{\rm crit}^* \exp\left[\frac{\mu_c L}{R} \left(\frac{1}{T_{\rm crit}^*} - \frac{1}{T} \right) \right]$, where $L$ is the latent heat chosen to be $1.85 \times 10^{10}$erg/g for that illustration, corresponding to the average value between the melting temperature and the critical endpoint temperature. \\

\begin{figure}[h!]
\vspace{10pt}
\centering
\includegraphics[width=\linewidth]{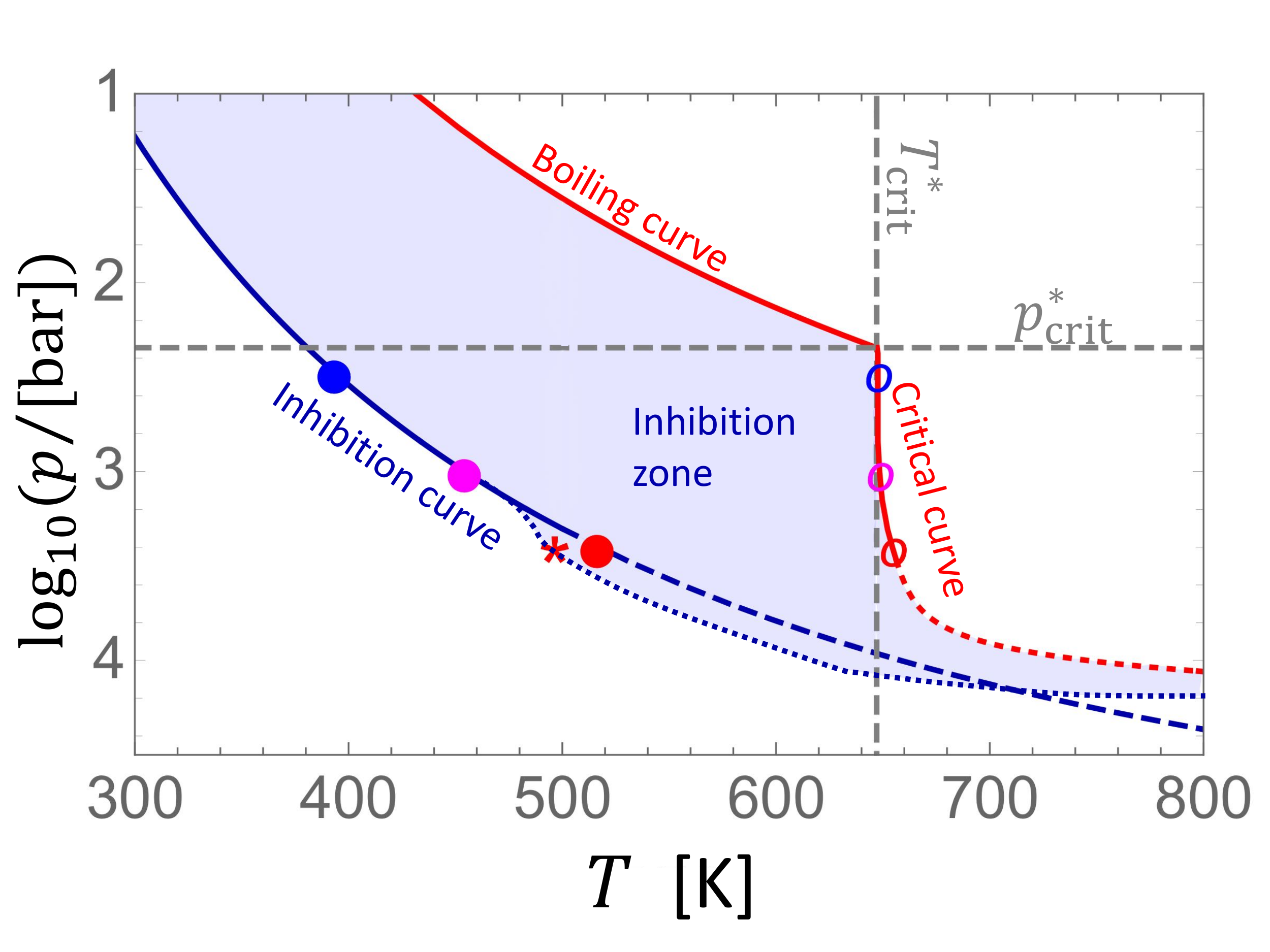}
\caption{Temperature at the boundaries of the stable layer as a function of envelope basal pressure, using water. 
The blue and red curves correspond to the top and bottom of the stable layer respectively.  
As in Fig.~\ref{fig:intersection}, we use solid curves where we have data and dashed curves for the ideal gas approximation. 
We further use a dotted curve for the interpolated critical curve and for a notional (non-quantitative) altered inhibition curve that accounts for the non-ideal coexistence behavior of the water--hydrogen system. 
The intersection marker symbols from Fig.~\ref{fig:intersection} (the blue, magenta, and red filed and open circles and the red star) are shown here for context. 
}
\vspace{10pt}
\label{fig:inhibition_zone}
\end{figure}

\begin{figure}[h!]
\vspace{10pt}
\centering
\includegraphics[width=\linewidth]{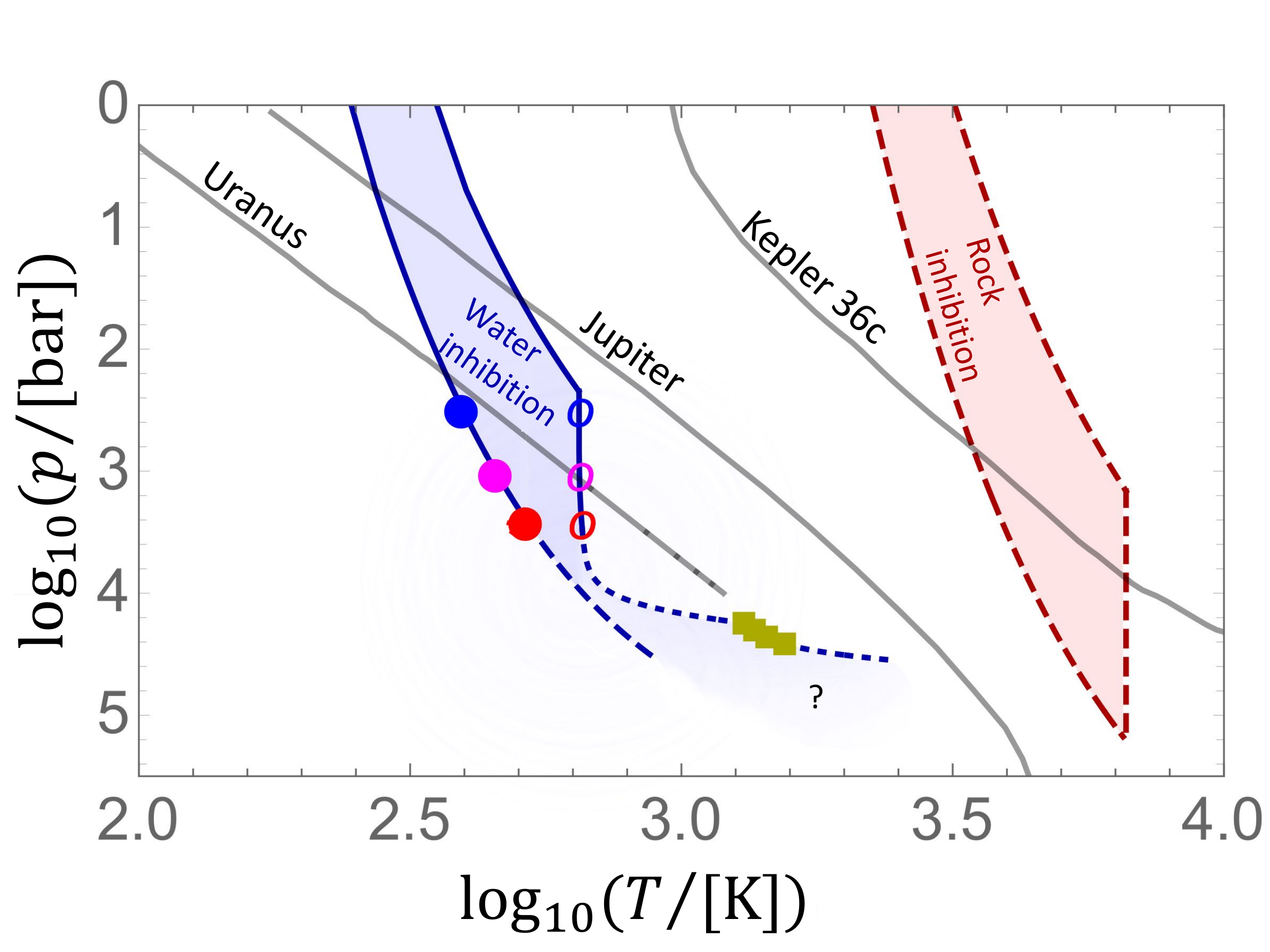}
\caption{Examples of pressure-temperature profiles for planets in our Solar System and galaxy. 
If the planet mass is dominated by an ocean (infinite reservoir) with a low-mass hydrogen envelope, convection will be inhibited in the shaded region, due to water in the blue region and silicates in the red region. 
While the red region uses the ideal gas approximation, the blue region extends into higher pressure regimes; we use a color gradient and question mark to emphasize the behavioral uncertainty under these conditions. 
Data for Uranus are extrapolated from \cite{lindal+1987}, the Jupiter model is from \cite{hubbard-militzer2016}, and the Kepler 36c model shortly after isolation from \cite{bodenheimer+2018}.
Colored points correspond to the same symbols from Figs.~\ref{fig:intersection}~and~\ref{fig:inhibition_zone}, with added yellow square points corresponding to data from \cite{bali+2013}.}
\vspace{10pt}
\label{fig:examples}
\end{figure}

This critical curve in Figures~\ref{fig:inhibition_zone}~and~\ref{fig:examples} interpolates between data from \cite{seward-franck1989} and \cite{bali+2013}. 
The shaded inhibition zone in Fig.~\ref{fig:inhibition_zone} corresponds to the equivalent shaded areas in Figures~\ref{fig:examples}~and~\ref{fig:Tps}. 
We shade this particular region for convenience, because the behavior of the inhibition and critical curves are unknown at higher temperatures and pressures, although the reader should understand that in reality the inhibition zone should extend everywhere between the inhibition curve and the critical curve. 
This is a key point that warrants emphasis; in Sect.~\ref{sec:evol} we use the inhibition zone as shown in red in Fig.~\ref{fig:examples}. 
That shaded zone bears a sharp point at the bottom, at which the inhibition temperature approaches the critical endpoint temperature. 
However in reality, the critical temperature depends on composition, we actually expect the critical temperature to curve strongly toward higher temperatures as shown in Fig.~\ref{fig:inhibition_zone}. 

We expect the basic thermodynamics to play out in much the same way for a mixture of silicates and hydrogen. 
We show roughly where we expect the silicate inhibition zone to be in Fig.~\ref{fig:examples}, and will use this inhibition zone again in Fig.~\ref{fig:Tps}. 
For these figures we use the ideal gas approximation for silicates, with the saturation vapor pressure of pure silicates taken from ab initio simulations \citep{seward-franck1989}. 
We use $\mu_B=20$ for silicates, corresponding to the mean value for an atomic disaggregation of $SiO_2$, while we use $\mu_A \sim 1.27$ corresponding to a mixture of atomic hydrogen and helium with a 10\% helium number fraction. 
Different choices for these values of $\mu_B$ and $\mu_A$ affect the placement of the inhibition curve by changing $\epsilon \equiv \mu_B / \mu_A$; from Eq.~\ref{eq:criterion_simple} one can see that increasing $\epsilon$ tends to decrease $x_{\rm inh}$ and vice versa.
 \\

Although we use the ideal gas approximation to delineate the rock inhibition zone in Figures~\ref{fig:examples} and \ref{fig:Tps}, in reality we expect 
the coexistence curves for the silicate--hydrogen system to look qualitatively similar to the water--hydrogen system, albeit with both pressure and temperatures increased by about an order of magnitude owing to the order-of-magnitude larger bond energies within silicate atoms compared to weaker hydrogen bonding between water molecules. 
Therefore, we expect the mixing ratio beneath the stable layer to depend on the pressure formation conditions. 
The mixing ratio immediately beneath the stable layer is set by the intersection of the solid black and colored curves in Fig.~\ref{fig:intersection}, marked with open circles. 
One should take note that the critical curve veers strongly to the left in Fig.~\ref{fig:intersection}, so that the behavior of the critical curve is very important when $p \gg p_{\rm crit}^*$. 
Fig.~\ref{fig:intersection} is a major finding of this work, as it demonstrates that even with an unlimited abundance of condensates, a stable layer due to convective inhibition will still have well defined boundaries. 
Furthermore, in general our model permits hydrogen to exist below the stable layer. 
We discuss hydrogen pollution of the core in the conceptual sense in more detail in Sect.~\ref{core-pollution}. 
\\

We must also consider chemistry particular to silicates that does not apply to ices that condense at temperatures cool enough that individual molecules do not dissociate. 
For silicates, unlike water, the atomic composition of the condensate in the condensed phase will not in general be identical to the atomic composition of the vapor phase \citep{xiao-stixrude}. 
Furthermore, unlike a mixture of hydrogen and water, hydrogen can react chemically with silicates, complicating the approximation of pure substances in coexisting phases \citep[e.g.,][]{fegley}. 
The chemistry is varied and complex, but likely one example is the equilibrium between silicate vapor and silane gas, for example,
\begin{equation}
\label{eq:silane}
SiO_{2 ~(l)} + 4 H_{2 ~(g)} \rightleftharpoons 2H_2O_{(g)} + SiH_{4 ~(g)}.
\end{equation}
Detailed modeling of the complicated series of hundreds of simultaneous equilibrium chemical reactions is beyond the scope of this work \citep[but has been modeled in prior works; see e.g.,][]{schaefer-fegley2009}, though we note its effect with respect to convective inhibition can still be accurately modeled using the generalized approach from Sect.~\ref{sec:proof}.\footnote{The derivation would follow the same steps, but one must use different assumptions and the final right-hand side term in Eqs.~\ref{eq:effective2} and \ref{criterion} would take a different form depending on the stoichiometry of the chemical reaction. 
}
\\

Having now discussed in detail the complications arising from thermodynamic considerations of critical mixtures primarily using existing data for water to illustrate, we now will transition to discussing practical consequences for super-Earth internal structure and evolution in the following Sect.s. 
As shown in Fig.~\ref{fig:examples}, Kepler 36c is an example of a planet whose early temperature structure intersects the rock inhibition zone, indicating that subsequent thermal evolution will be affected by convective inhibition. 
To attempt to quantify the effect convective inhibition will have on subsequent thermal evolution, we use the ideal gas approximations for convenience, with the inhibition zones demarcated by the red shaded region of Fig.~\ref{fig:examples}. 
The boundaries of this region, the ideal gas inhibition curve, the ideal gas boiling curve, and the critical endpoint temperature, are chosen for convenience because they are values for which we have preliminary data for silicates \citep{xiao-stixrude}. 
However, we hope in this section we have impressed upon the reader that these approximations are incomplete when considering a critical mixture, and our results should be regarded as notional orders of magnitude rather than quantitatively exact. 
Nevertheless, even without more precise data, there are already an abundance of impactful consequences of convective inhibition with an ocean.

\section{Structure}
\label{sec:structure}

Having established the theoretical foundations for convective inhibition by silicate vapor, we now seek to delineate under what circumstances these considerations are relevant, and to comment on the potential effects convective inhibition will have on super-Earth internal structure. 
In Sect.~\ref{sec:when} we outline a simple envelope model. 
We find for a planet in a state of convective inhibition, that the mass of the envelope above the stable layer together with its equilibrium temperature uniquely defines its structure and luminosity. 
In Sect.~\ref{stable-layer} we place limits on the thickness of the stable layer. 
Finally, in Sect.~\ref{sec:below} we speculate on the planet's properties below the stable layer, and argue for why a more precise understanding must be predicated on so far unavailable data. 

\subsection{Envelope structure}
\label{sec:when}
We consider the envelope structure under the following approximations: neglecting self-gravity of the atmosphere, an ideal gas equation of state, an adiabatic temperature gradient below the radiative-convective boundary (RCB), and a plane-parallel geometry. 
We further use the approximations for the coexistence of silicate vapor with hydrogen gas outlined in the previous Sect.. 
Under the cold Murnahan equation of state, the radius of a silicate core scales as the quarter power of its mass \citep{seager+2007}
\begin{equation}
\label{eq:murnahan}
R_c=R_\oplus (M_c/M_\oplus)^{1/4}.
\end{equation} \\

We model the saturation vapor pressure as
\begin{equation}
p_s = \exp(A-B/T),
\end{equation}
where $p_s$ is the saturation vapor pressure in GPa, and the parameters $A$ and $B$ are 11.8 and 45,000 respectively, fitting ab initio simulations from \cite{xiao-stixrude}. 
Convective inhibition is initiated at pressure $p_1$ and temperature $T_1$ satisfying 
\begin{equation}
\label{eq:p1T1}
\frac{p_s(T_1)}{p_1} = x_{\rm inh}(T_1).
\end{equation}

Under our approximations, the total pressure at the bottom of the envelope (or equivalently the top of the core) is $p_c = \frac{M_e g}{4\pi R_c^2}$ where $M_e$ is the total mass in the envelope. 
Conveniently, because $R_c \propto M_c^{1/4}$ (Eq.~\ref{eq:murnahan}), this can be rewritten to be independent of core mass and radius: 
\begin{equation}
p_c = \frac{G M_e M_\oplus}{4\pi R_\oplus^4}.
\end{equation} 
The total pressure at the core includes both gas mass $M_{XY}$ and some mass from silicate vapor, which can be non-negligible. 
The total mass of the envelope, then, is 
\begin{equation}
\label{eq:envelope-mass}
M_e = M_{XY} + 4\pi R_\oplus^4 \int_{0}^{p_c} \frac{\epsilon p_s(T(p))}{p G M_\oplus} dp,
\end{equation}
where $\epsilon$ is the molecular mass ratio of volatile vapor to dry air. 
In the following calculations we assume the stratosphere is cold enough that it contains negligible silicate vapor such that the lower bound of the right-hand side integrand of Eq.~\ref{eq:envelope-mass} becomes $p_{\rm{rcb}}$ instead of $p=0$. 
We also assume $p_1 \sim p_c$ (see Sect.~\ref{stable-layer}). 
The above equation is independent of core mass. 
Therefore for a given envelope mass, the pressure at the top of the core is uniquely determined. \\

\begin{figure}[h!]
\vspace{10pt}
\centering
\includegraphics[width=\linewidth]{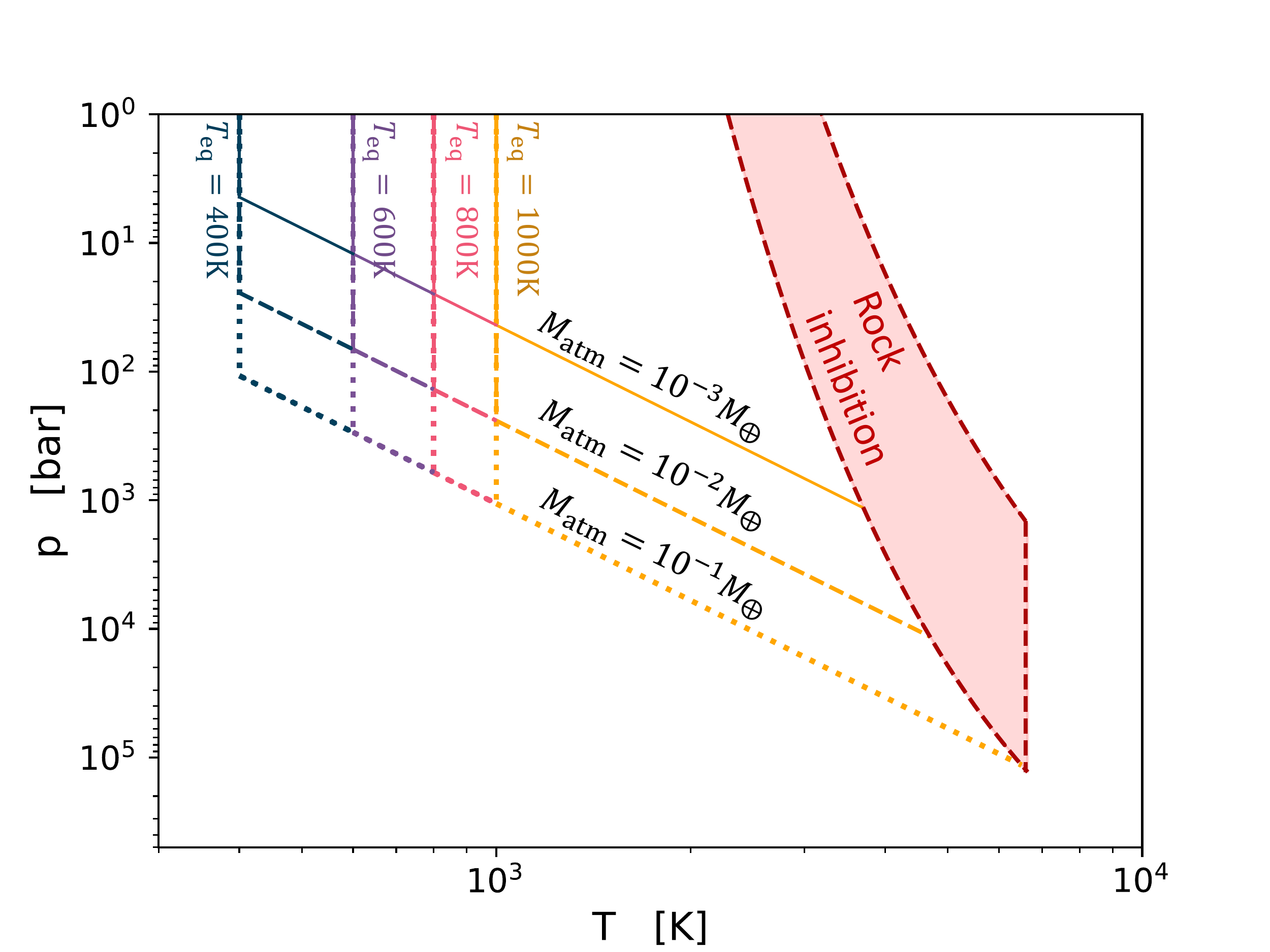}
\caption{Envelope pressure--temperature profiles for a sample of envelope properties. 
The blue and red curves represent the temperature at the top and bottom of the stable layer, respectively; for a saturated gas mixture, convection is shut off between the blue and red curves. 
In the absence of data, this figure assumes $x_s = p_s / p$ and does not consider the temperature dependence of latent heat or the compositional dependence of the critical curve. 
See Fig.~\ref{fig:sketch} for a conceptual sketch of the temperature structure below the envelope.
}
\vspace{10pt}
\label{fig:Tps}
\end{figure}

\begin{figure}[h!]
\vspace{10pt}
\centering
\includegraphics[width=\linewidth]{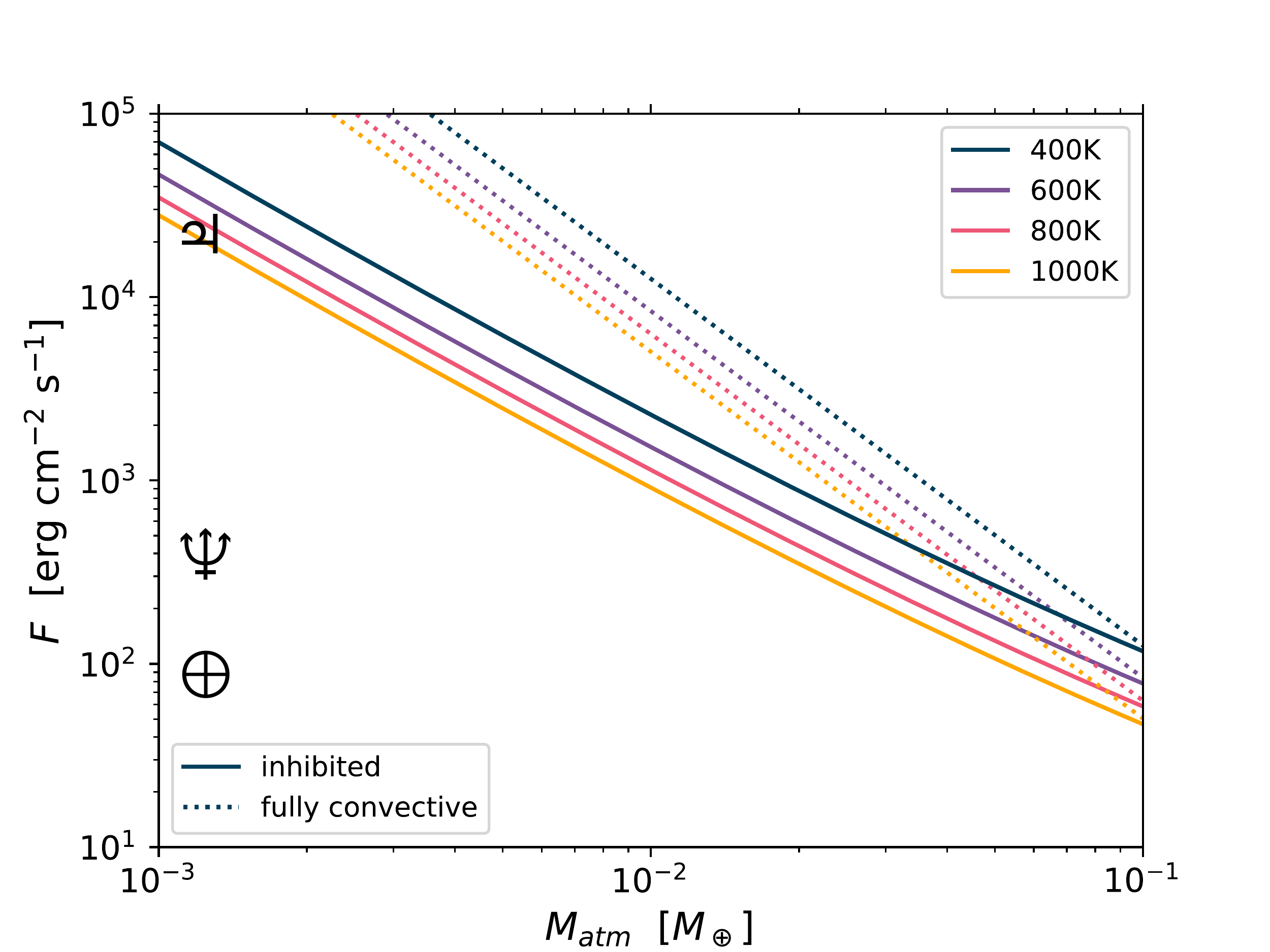}
\caption{Core-envelope heat flux as a function of envelope mass. 
Temperatures in the legend correspond to $T_{\rm eq}$. 
Symbols on the left of the plot show the contemporary internal heat fluxes of Earth, Neptune, and Jupiter for comparison. }
\vspace{10pt}
\label{fig:fluxes}
\end{figure}

Because $T_1$ is uniquely determined for a given $p_1$ according to Eq.~\ref{eq:p1T1}, this sets the adiabat for the convective part of the envelope. 
Highly irradiated super-Earths are expected to have effective temperatures very near their equilibrium temperatures \citep[e.g.,][]{ginzburg+2016}, and the temperature at the RCB will be near the effective temperature. 
Using the adiabatic relationship 
\begin{equation}
\label{eq:ab}
T_1 = T_{eq} \left(\frac{p_1}{p_{rcb}} \right)^{\nabla_{ab}},
\end{equation}
where $\nabla_{ab} \equiv \frac{d\ln T}{d\ln p}|_{ab}$ is the adiabatic temperature gradient $\nabla_{ab} = \frac{\gamma-1}{\gamma}$, which is invariant if the Gr\"{u}neisen parameter is invariant. 
Following \cite{owen-wu2017}, we take $\gamma$ to be a constant 5/3. 
By combining these considerations, we can uniquely determine the RCB pressure by specifying the total mass $M_{XY}$ of gas in the envelope. 
The intrinsic luminosity scales with the depth of the RCB \citep{ginzburg+2016}
\begin{equation}
\label{eq:luminosity}
L = \frac{64 \pi \sigma T_{\rm rcb}^4 R_B'}{3 (\rho \kappa)_{\rm rcb}},
\end{equation}
where $R_B'$ is the so-called ``modified Bondi radius,'' 
\begin{equation}
R_B' \equiv \frac{\gamma - 1}{\gamma} \frac{G M_c \mu_d}{k_B T_{\rm rcb}}.
\end{equation}
Eq.~\ref{eq:luminosity} intuitively states that the planetary flux is inversely proportional to the optical depth of the RCB, i.e., $F \sim \sigma T_{\rm eq}^4/\tau_{\rm rcb}$, and is valid for large $\tau$. 
Following \cite{freedman+2008}, we assume the opacity scales linearly with pressure, so that the planetary flux diminishes rapidly with increasing depth of the radiative-convective boundary. 
We plot the planetary flux as a function of envelope mass in Fig.~\ref{fig:fluxes}. 
Fig.~\ref{fig:fluxes} compares two assumptions for atmospheric models. 
In both cases, the basal pressure is determined by the envelope mass. 
In the inhibited case, the corresponding basal temperature is $T_1$ from Eq.~\ref{eq:p1T1}. 
In the uninhibited case the basal temperature is $T_{\rm{crit}}^*$. 
These values are chosen so that the heat of the core is the same in both models.
\\

The reader may notice that the $10^{-1} M_\oplus$ curve in Fig.~\ref{fig:Tps} intersects the inhibition curve very near the sharp point at the bottom of the labeled inhibition zone. 
That is the reason in the abstract we say that our results pertain to planets with envelope masses up to 10\%$M_\oplus$, because at higher pressures the adiabat does not intersect the labeled inhibition zone.  
However, we recall from Sect.~\ref{sec:theory} that the sharp point is in fact an artifact of our approximations, and that in reality we expect the critical curve to depart from the critical endpoint temperature, so that convective inhibition may still relevant to larger envelope masses. 
This approximation is also related to the fact that the uninhibited model flux approaches the inhibited model flux around $10^{-1} M_\oplus$ in Fig.~\ref{fig:fluxes}. 
Again, this is an artifact of our approximations, and the behavior would be different with a more realistic model for the inhibition and critical curves (see Fig.~\ref{fig:inhibition_zone}). \\

Fig.~\ref{fig:fluxes} is centered around an envelope mass of $10^{-2}M_\oplus$. 
We note for clarity that this value is in Earth masses, not as a fraction of the core mass. 
So, for example, a $5M_\oplus$ super-Earth that is 1\% gas by mass should be read as $5 \times 10^{-2}$ on Figures~\ref{fig:fluxes} and \ref{fig:tcools}. 
We further note that an atmospheric mass of $\sim 10^{-2} M_c$ is a realistic estimate for super-Earths based on hydrostatic equilibrium between the core and the gas disk during formation \citep[see, e.g.,][]{ginzburg+2016}. \\

The behavior of Fig.~\ref{fig:fluxes} shows a monotonically decreasing relationship for planetary flux as a function of envelope mass. 
Intuitively, one can understand the behavior this way: 
the pressure at the bottom of the envelope scales linearly with envelope mass. 
Under the thin stable layer approximation, we can then use the relationship from Equation \ref{criterion} to uniquely determine the temperature $T_1$ at the bottom of envelope such that $x\rightarrow x_{crit}$ (see Fig.~\ref{fig:Tps}). 
We note that in Fig.~\ref{fig:Tps}, the thermodynamic conditions at the base of the envelope, ($p_1$, $T_1$), depend on envelope mass only. 
Higher equilibrium temperature planets have deeper RCBs.
Given the effective temperature, we can use the adiabatic relationship to place the RCB. 
Assuming fixed $T_1$, we can then use Eq.~\ref{eq:luminosity} to estimate that the luminosity (or flux, for fixed radius) should scale approximately as the inverse square of the envelope mass. 
In reality this is an overestimate; accounting for the fact that $T_1$ will increase as $p_1$ increases accounts for the somewhat shallower dependence of flux on envelope mass. \\

Fig.~\ref{fig:fluxes} also shows that the flux of the planet is not too sensitive to the planet's effective temperature, and that the flux decreases with increasing effective temperature. 
From Eq.~\ref{eq:luminosity}, we see that $F \propto T_{\rm eq}^4 / p_{\rm rcb}^2$, where the flux is the core-envelope internal heat flux (this proportionality holds when $\nabla_{\rm ab} \sim 1/4$). 
For fixed $M_{atm} \implies (p_1$, $T_1)$ will likewise be fixed.  
Therefore we can infer the expected scaling for $p_{rcb}$ from Eq.~\ref{eq:ab}. 
Therefore, by combining Eqs.~\ref{eq:luminosity} and \ref{eq:ab} using $\nabla_{\rm ab} = 1/4$, 
we then find the scaling relationship for flux at fixed envelope mass to be $F \sim T_e^{-1}$. 
This result is consistent with Fig.~\ref{fig:fluxes} that shows internal heat flux decreasing with increasing equilibrium temperature. 
We note that this result is sensitive to the adiabatic lapse rate, set by the Gr\"{u}neisen parameter. \\

The critical pressure of silicates is about 1.4kbar \citep{xiao-stixrude}, corresponding to an envelope mass of about $10^{-3}M_\oplus$. 
For much lower mass, the flux is sufficiently high that the core can cool quickly, and prior models subsequently work to a good approximation. 
Therefore for all cases of interest to us, the core will initially be in a supercritical state, rather than a magma ocean. \\

\subsection{Estimating the thickness of the stable layer}
\label{stable-layer}
We must challenge our assumption that the stable layer can be approximated as infinitesimally thin. 
Furthermore, we will challenge our assumption that the layers of the planet behave distinctly --- either fully convective or fully stable --- given the intrusion of eddy diffusivity.  
The thermal transport properties of the materials of interest under the relevant thermodynamic conditions are poorly constrained, but we can set reasonable estimates on their order of magnitude. 
We consider three thermal transport mechanisms of interest: conductive, radiative, and advective. \\

In the absence of thermal transport data at high temperatures and pressures, we resort to the ideal gas estimate $k \sim \rho \lambda c_v \sqrt{\frac{2k_B T}{\pi m}}$. 
Thermal conductivity should be roughly independent of density, and the mean free path $\lambda$ is inversely proportional to density. 
Using characteristic numbers, and using the critical temperature of silicates as an upper bound, we find the thermal conductivity to be $k \sim 10^5$~g~cm~s$^{-3}$K$^{-1}$. \\

Radiative heat transport in the limit where the mean free path is small compared the length scales of the problem (valid for high pressures), we can estimate the equivalent thermal conductivity due to radiative transfer as $k \sim \frac{4 \sigma T^3}{\rho \kappa}$. 
The relevant parameters are reasonably well-constrained in this context, except the opacity. 
The thermodynamic conditions of interest are thousands of kelvins, set by silicate vapor pressure curves, and kilobars to hundreds of kilobars, set by the mass of the envelope. 
The high pressures imply a large amount of collisional opacity of hydrogen, while the temperatures are so high that partial ionization and highly opaque free electrons are expected. 
Furthermore in our case the stable layer is a mixture of hydrogen and silicates, so that the stable layer should be a cosmic mixture of many elements, not dominated by hydrogen or any other single substance that may have transparent windows in its transmission spectrum. 
We therefore expect this environment to be highly opaque. 
Following \cite{freedman+2008}, the opacity for 1 kbar to hundreds of kbar (the pressure range of interest) should be significant, plausibly on the order of 1-100cm$^2$g$^{-1}$. 
To be cautious we assume a low opacity on the order of 1cm$^2$g$^{-1}$. 
In this case we obtain a characteristic thermal conductivity on the order of $10^7$g~cm~s$^{-3}$~K$^{-1}$. 
If the material is more opaque, the corresponding equivalent thermal conductivity will be smaller, plausibly comparable to the microscopic thermal conductivity $\sim 10^5$g~cm~s$^{-3}$~K$^{-1}$. 
Our upper estimate is two orders of magnitude larger than the estimate for conduction, and would therefore dominate thermal transport. \\

Finally, we consider advective heat transport. 
Although the stable layer is called as such because it is stable against ordinary large-scale overturning convection, we cannot assume this layer is completely stagnant. 
We know statically stable layers of the Earth's atmosphere can nevertheless involve considerable eddy diffusivity due to, for example, breaking gravity waves \citep{dornbrack1998, garcia-solomon1985}. 
We can attempt to constrain the order of magnitude of this process, although it is highly uncertain. 
In order to move material upward, one must do work against gravity. 
We can assume as a first approximation that the eddy diffusivity behaves like thermal diffusivity, but acts on deviations away from adiabaticity. 
For heat flux on the order of $10^3$erg~cm$^2$~s$^{-1}$ (see Fig.~\ref{fig:fluxes}), this predicts an eddy diffusivity of comparable order-of-magnitude to ordinary thermal diffusivity. 
Accounting for all these considerations, it appears radiative heat transport is most likely to dominate. \\

Using our rough upper bound for equivalent thermal conductivity of the system, we can estimate the thickness of the stable layer. 
From Fourier's law, $F=-k \nabla T$, or $L \sim k \Delta T / F$. 
For a 1\%$M_\oplus$ envelope, we estimate that $F$ should be on the order of $10^3$erg~cm$^{-2}$~s$^{-1}$, and use a characteristic value for the temperature difference $\Delta T \sim 10^3$K (see Fig.~\ref{fig:Tps}). 
Therefore the thickness of the stable layer should be on the order of 100km, or about $<$1\% the expected radius of a super-Earth. 
The corresponding pressure drop would be on the order of a kbar, about 10\% the pressure overlying the stable layer for a 1\%~$M_\oplus$ envelope. 
For smaller atmospheric masses, the thermal flux will be greater and the density/opacity smaller, while for larger atmospheric masses the thermal flux can be smaller but the density/opacity larger. 
Therefore for our conservative estimates for relatively efficient radiative heat transport, the stable layer can be non-negligible in thickness and mass content, and this will add a small but non-negligible correction to Figures~\ref{fig:fluxes} and \ref{fig:tcools}. \\

\subsection{Below the envelope}
\label{sec:below}
\begin{figure}[h!]
\vspace{10pt}
\centering
\includegraphics[width=\linewidth]{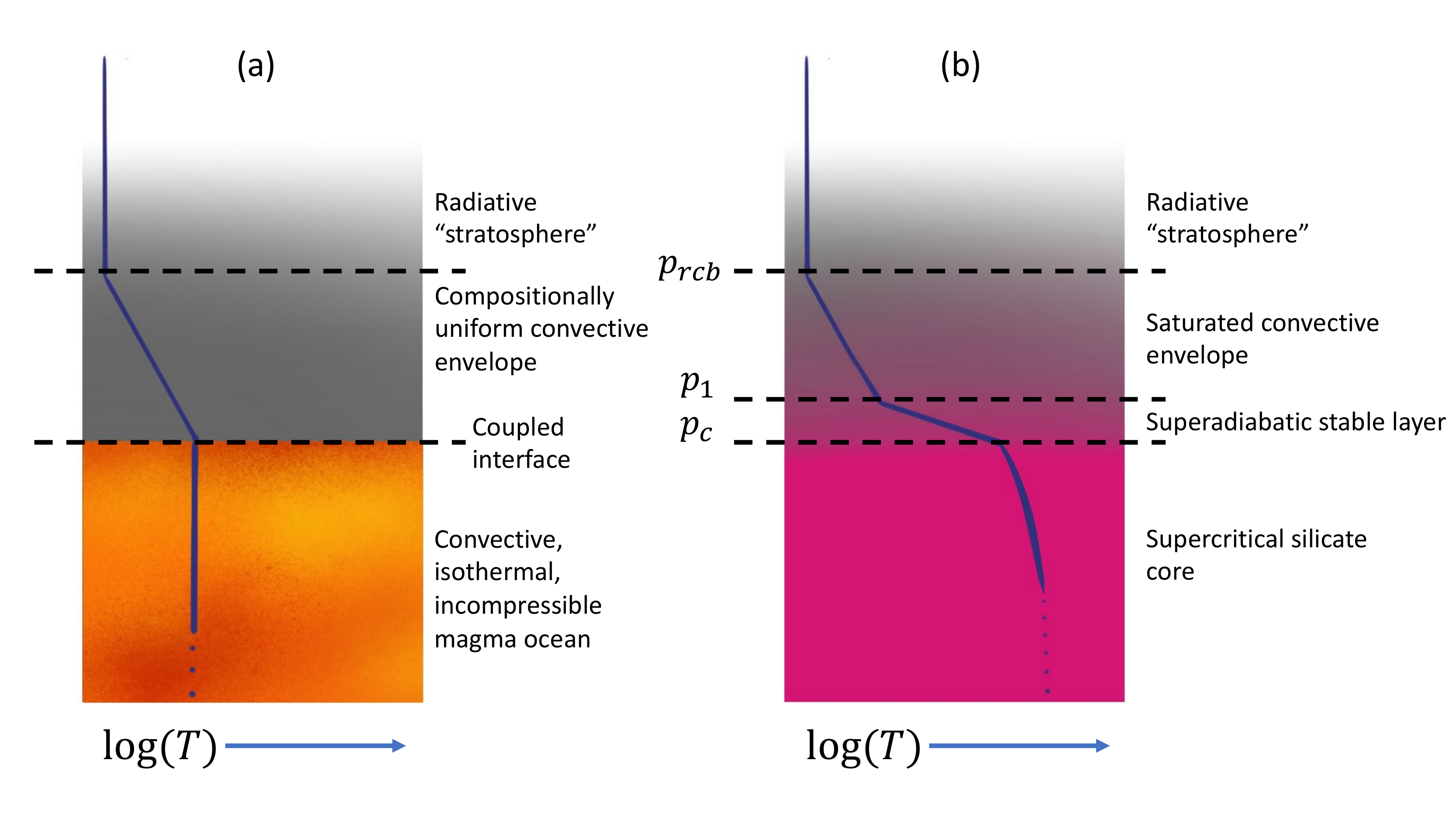}
\caption{Sketch of the proposed temperature structures of silicate planets with gaseous envelopes described in Fig.~\ref{fig:cartoon}. 
See Fig.~\ref{fig:Tps} for explicitly quantified envelope pressure-temperature diagrams; this figure is only intended to be an illustrative sketch.
}
\vspace{10pt}
\label{fig:sketch}
\end{figure}

We will keep our discussion of the planetary structure beneath the envelope largely conceptual and brief, as this preliminary exploratory work primarily focuses on the part of the planet above the stable layer ($p_1$ is Fig.~\ref{fig:sketch}).  
The envelope structure we propose here is very similar to prior models \citep[e.g.,][]{ginzburg+2016}. 
However, we will comment here on the structural assumptions that underpin Sect.~\ref{sec:evol}, as well as comment on the shortcomings of these assumptions to be further addressed in Sect.~\ref{sec:followup}. \\

The bottom of the stable layer truncates at the critical curve (see Fig.~\ref{fig:intersection}). 
For the purposes of Sect.~\ref{sec:evol}, we refer to everything below the stable layer as the ``core,'' although this terminology may be misleading. 
Thermodynamics permits the layer beneath the stable layer to be highly enriched in hydrogen. 
The exact enrichment depends on formation conditions and the details of the coexistence curve under the relevant conditions, and there may be additional compositional gradients and boundaries \citep[cf.][]{bodenheimer+2018, brouwers-ormel2020}. 
Furthermore, the behavior of the core at depth depends on currently unavailable data about the physical properties of the supercritical rock/gas mixture under the relevant conditions. 
For example, the relative steepness of the critical curve and the adiabat can have profound implications for the expected interior structure \citep[cf.][]{bailey-stevenson2021}. 
We therefore refrain from advancing a detailed core model, which would be premature without the requisite data. \\

Nevertheless, we must employ some model. 
Having established the impossibility of an accurate model, we therefore opt for a simple model. 
Throughout this work, we have treated the core radius as scaling like the cold Murnahan equation of state. 
As the name implies, this equation of state requires corrections for a hot, supercritical core. 
However, as we argued in Sect.~\ref{sec:theory}, there are currently too many unknowns to attempt a serious dynamic core model, as the results would depend on too many unsubstantiated assumptions about the behavior of critical mixtures of rocks and gas. 
Therefore, in Sect.~\ref{sec:evol} we model the core as a well mixed, fully convective incompressible magma.
We apply this treatment as an end-member case for the purpose of placing a constraint on the orders-of-magnitude of the timescale.  
However, in reality the core is likely to be supercritical and significant quantities of gas may pollute the core. 
Sect.~\ref{core-pollution} qualitatively analyzes the possible effect of such core pollution on our results. \\

In our model we consider cases where the heat flow is fully balanced, that is, radiative-convective quasi-equilibrium. 
Therefore, the heat flow across the stable layer is balanced by heat flow into space. 
Because we regard the stable layer to be thin ($p_1 = p_c$), then, the atmospheric structure remains static until the core cools to below $T_1$, at which point convective inhibition ceases and convective thermal contact between the envelope and the core asserts itself. 
Subsequent cooling then progresses as prior theories \cite[e.g.,][]{ginzburg+2016} suggest. 
Now we must compute the timescale on which the core cools from an initially hot ($T \sim T_{\rm crit}$) state down to the temperature at the top of the stable layer.

\section{Evolution}
\label{sec:evol}
\begin{figure}[h!]
\vspace{10pt}
\centering
\includegraphics[width=\linewidth]{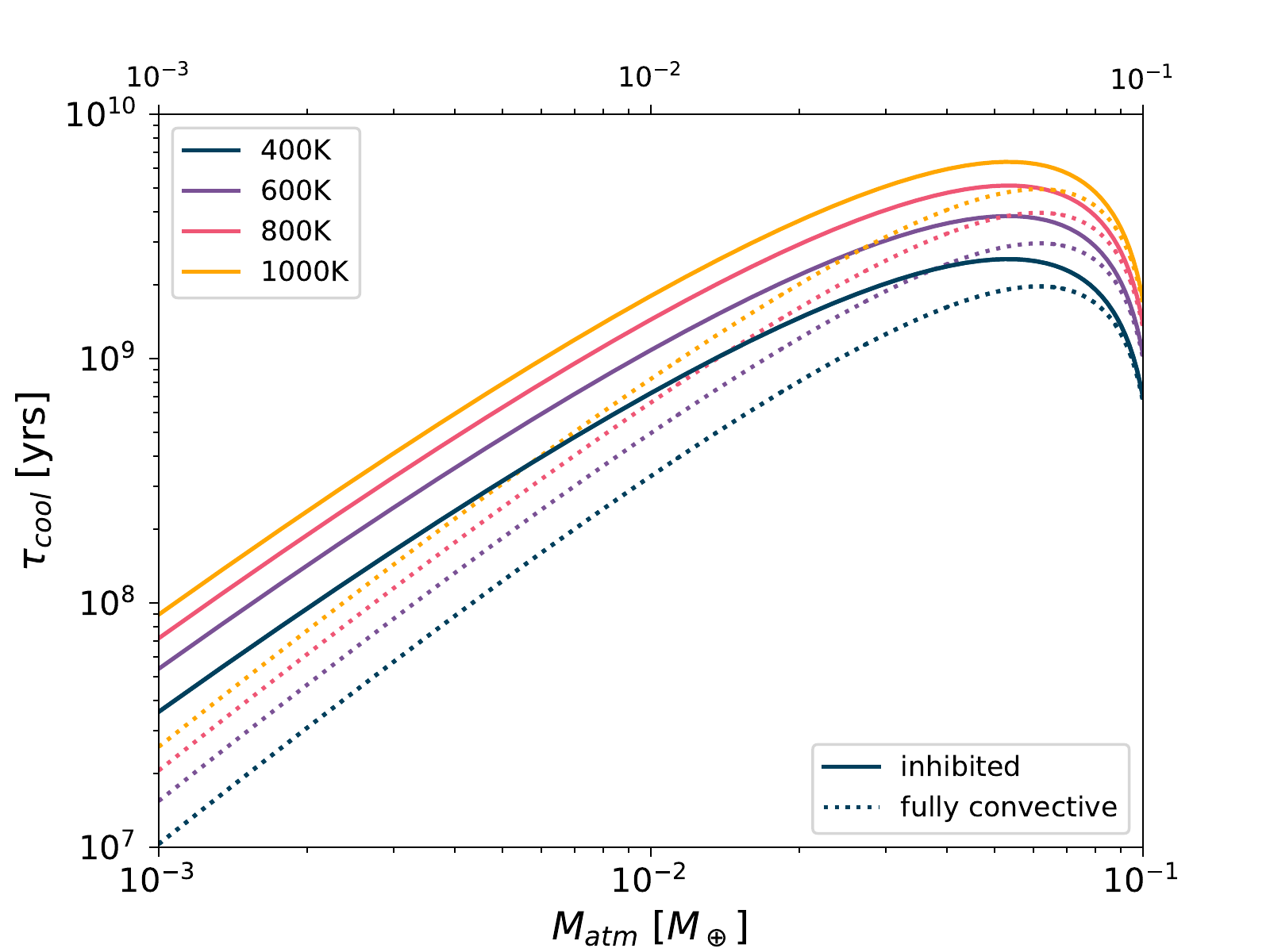}
\caption{Cooling timescales for a super-Earth computed using Eq.~\ref{eq:tcool}. 
The dotted curves show the cooling timescale without convective inhibition with the core temperature at $T_{\rm crit}^*$. 
The solid curves include convective inhibition as described in this work. 
}
\vspace{10pt}
\label{fig:tcools}
\end{figure}

Previous works \citep[e.g.,][]{vazan+2018} have identified that the existence of a gas envelope can considerably slow the thermal evolution process, such that a magma ocean evolutionary phase may persist for billions of years. 
Our findings here indicate that thermal evolution of the core could take place even more slowly, owing to the inefficient thermal transport through a stable layer. 
We begin by computing the cooling timescale of a body beginning in the state described in the previous Sect., with the top of core at the critical temperature and the bottom of the envelope at $T_1$ such that convection is inhibited. 
We assume in the radiative-convective quasi-equilibrium state that the envelope temperature profile is frozen, that is does not change in time. 
This occurs because, if the luminosity of the planet exceeds the heat transport from the core the envelope, then the convective envelope will cool, further thinning the stable layer, until the stable layer is sufficiently thin that the heat flow through the stable layer balances the heat flow out of the planet. 
In this section we argue that even if the planetary luminosity initially exceeds the heat flow between the core and envelope, the envelope will quickly relax into the equilibrium state. 
In radiative-convective quasi-equilibrium, 
\begin{equation}
L_{core}= - M_{core} c_v \frac{d T}{dt}.
\end{equation}
The cooling timescale that describes how long the planet remains in this stabilized state, then, is 
\begin{equation}
t_{cool} = \frac{M_{core} c_v (T_{\rm crit} - T_1)}{L_{\rm core}}.
\label{eq:tcool}
\end{equation}
We assume the heat capacity of the core agrees with the ideal gas limit \citep{bolmatov+2013}, consistent with prior works \citep[e.g.,][]{ginzburg+2016}. 
Assuming the luminosities computed in Fig.~\ref{fig:fluxes}, we compute cooling timescales shown in Fig.~\ref{fig:tcools} assuming a 5$M_\oplus$ core. 
The cooling timescale for the solid curves in Fig.~\ref{fig:tcools} use Eq.~\ref{eq:tcool}. 
However, when considering a planet cooling without convective inhibition (dotted curves), we must account for the non-constant luminosity, as the luminosity of the planet will diminish as the atmosphere and core cool together. 
We can compute the core cooling time analytically by integrating Eq.~\ref{eq:tcool} while accounting for a non-constant luminosity. 
We can rewrite the luminosity from Eq.~\ref{eq:luminosity} as 
\begin{equation}
\label{eq:luminosity2}
L_{\rm core} = \frac{128 \pi}{3} \sigma T_e^4 \frac{G M_c \nabla_{\rm ab} p_0}{p_{\rm rcb}^2 \kappa_0}.
\end{equation}
As the atmosphere cools, the RCB becomes deeper. 
Under our assumptions, the relationship between $p_{\rm rcb}$ and the temperature at the bottom of the atmosphere $T_{\rm btm}$ is 
\begin{equation}
p_{\rm rcb} = \frac{G M_\oplus M_{\rm atm}}{R_\oplus^4} \left(\frac{T_e}{T_{\rm btm}} \right)^{1/\nabla{\rm ab}}.
\end{equation}
Therefore we have a relationship between the time-dependent top of core temperature, and the luminosity. 
We know the atmospheric state when $T_{\rm btm} = T_{\rm crit}^*$ and at the end state when $T_{\rm btm} = T_1$. 
Therefore we can integration Eq.~\ref{eq:tcool} by substitution to obtain 
\begin{equation}
\label{eq:uninhibited}
\Delta t_{\rm uninhibited} = \frac{3 c_v \kappa_0 G M_\oplus^2 M_{\rm atm}^2}{128 \pi \sigma T_e^3 p_0 R_\oplus^8 (\nabla_{\rm ab} - 2)} \left[ \left( \frac{T_e}{T_1} \right)^{1/\nabla_{\rm ab}} - \left( \frac{T_e}{T_{\rm crit}^*} \right)^{1/\nabla_{\rm ab}} \right]
\end{equation}
We use this analytic expression for the dotted curve in Fig.~\ref{fig:tcools}.
\\

The upper abscissa in Fig.~\ref{fig:tcools} occurs because in this simplified model $T_1 \rightarrow T_{\rm crit}^*$ when $M_{\rm atm} \sim 10\% M_\oplus$ (the bottom of the atmosphere in Fig.~\ref{fig:Tps} is very close to the tip of the inhibition zone); therefore, the numerator in Eq.~\ref{eq:tcool} becomes smaller.  
We use this simplifying assumption due to the absence of data, but in reality the tip we see in the silicate inhibition zone in Figures~\ref{fig:examples} and \ref{fig:Tps} probably does not behave like that.
In reality, we expect the actual critical curve for a silicate-hydrogen mixture to increase at high temperatures (see Sect.~\ref{sec:theory}, Fig.~\ref{fig:inhibition_zone}). 
\\

The cooling times in Fig.~\ref{fig:tcools} refer to the time it takes the core to cool sufficiently that convective inhibition is no longer important. 
These timescales are somewhat longer than the cooling timescales for the uninhibited case, by about a factor of 5 for a 1\% $M_\oplus$ hydrogen envelope. 
The calculations for Fig.~\ref{fig:tcools} were performed assuming a 5$M_{\oplus}$ core, consistent with Fig.~\ref{fig:fluxes}, but the cooling timescale results are actually independent of core mass under our assumptions. 
From Eq.~\ref{eq:luminosity2}, the luminosity of core scales linearly with the mass of the core. 
Because the heat content of the core also scales linearly with its mass, the cooling timescale is unaffected by this choice. 
One can also see that our analytic expression for the cooling time in the uninhibited case, Eq.~\ref{eq:uninhibited}, is independent of $M_c$ for the same reason. 
This result is a consequence of our assumption for the equation of state of the core (Eq.~\ref{eq:murnahan}). 
If the mass radius scaling were to behave differently, the cooling timescale would no longer be independent of core mass. 
After the quantity we define as the cooling time has elapsed, the core and envelope can cool together, and one can subsequently employ existing evolution models \citep[e.g.,][]{ginzburg+2016, vazan+2018, chachan-stevenson2018}.
\\

The fact that $T_1 \rightarrow T_{\rm crit}^*$ for $p \gg p_{\rm crit}^*$ is an unphysical consequence of our treatment of the critical point as constant. 
A more accurate model must consider a composition-dependent critical curve, see Figs.~\ref{fig:intersection}~and~\ref{fig:examples}. 
A more accurate model requires coexistence data of critical mixtures, which will tend to increase the critical temperature and pressure compared to a pure substance. 
We therefore urge readers to focus more on the concepts here than the numbers, as more data are needed. 
\\

We must briefly inspect the assumption that we can treat the envelope as initialized in the equilibrium state. 
An envelope 1\% the mass of the core contains about 10\% of the molecules. 
Therefore if the envelope were warm at early time, then the envelope could cool to a lower luminosity state on timescales about an order-of-magnitude shorter than the time it would take to cool the core and envelope simultaneously. 
Furthermore, if the initial state of the envelope is higher entropy than our identified equilibrium state, then thermal transport from the core will be tremendously inefficient across the stably stratified envelope (we quantify the early heat transport efficiency  in the following paragraph). 
Therefore our estimates on cooldown timescales from Sect.~\ref{sec:evol} should be considered as lower bounds, because it may take even longer to cool if the envelope must first cool down to the equilibrium state. \\

At early times, the core will be hot and the internal heat flux of the planet high. 
We must therefore investigate the core cooling rate before the planet reaches global radiative-convective equilibrium. 
Assuming an initially adiabatic atmosphere, for example due to adiabatic compression when accreting isentropic disk gas, and if the envelope is initially saturated everywhere, then the bottleneck for core cooling will initially be the radiative heat loss due to an adiabatic gradient. 
The adiabatic temperature gradient under the ideal gas approximation for hydrogen will be $\nabla T \sim 10^{-5} (M_c / M_\oplus)^{1/2} K cm^{-1}$. 
Using our order-of-magnitude assumptions for opacity in the pressure regime of interest from Sect.~\ref{stable-layer}, we can place constraints on the early heat flow to be on the order of $10^0 - 10^2$ erg cm$^{-2}$ s$^{-1}$. 
Comparing these values to Fig.~\ref{fig:fluxes}, we see that the heat escape from the core at early times is much lower than the heat flux after achieving global radiative-convective equilibrium of the envelope, consistent with our assumptions.

\section{Future work}
\label{sec:followup}

In this Sect., we highlight natural consequences of this new model for super-Earth internal structure. 
Each of these considerations are fruitful topics for follow-up works, and provide additional motivation for the acquisition of further experimental data. 
In Sect.~\ref{core-pollution}, we discuss a major topic completely sidestepped in this work: the fact that supercritical hydrogen and silicates are expected to be miscible in all proportions. 
This may have important consequences for internal structure and evolution. 
Finally in Sect.~\ref{formation}, we comment on the importance of coupling our findings to a fully self-consistent and physics-based formation model. 
Each of these topics will be addressed in more detail in follow-up papers to this series.

\subsection{Hydrogen pollution in the core}
\label{core-pollution}
Recent formation models of super-Earths indicate a high metallicity region outside of the embryonic core polluted with hydrogen \citep{brouwers-ormel2020, bodenheimer+2018}. 
Above the critical temperature and pressure, silicates and hydrogen can mix in all proportions. 
The exact nature of the mixture depends on thermodynamics, and the formation scenario. 
During formation, the mixing ratio depends on the ambient thermodynamic conditions (Figures~\ref{fig:intersection}). 
At low temperatures, there will be two coexisting phases, one of which is nearly pure hydrogen-helium (some trace silicate vapor), while the other is nearly pure silicates (as droplets with trace dissolved hydrogen). 
Approaching the critical curve at high pressures, we expect the mixing ratio to approach 50\%. 
Indeed according to Fig.~\ref{fig:intersection} there is reason to think the stable layer should truncate at a temperature containing significant hydrogen enrichment. 
Therefore thermodynamically, significant hydrogen pollution of the core is allowed, but the amount and distribution depend on its formation conditions. 
The idea of hydrogen pollution into a magma core due to magma dissolution governed by Henry's law \citep{chachan-stevenson2018}, or due to a fugacity crisis \citep{kite+2019} has already been studied in detail; in this work we suggest hydrogen pollution can be even more efficient, because hydrogen and silicates can mix in all proportions as a supercritical mixture. 
\\

The effect of hydrogen in the core depends on its distribution and quantity, but the direction of the effect is clear: greater hydrogen enrichment will inflate the core size relative to pure rock. 
Hydrogen pollution can also provide a buffer against atmospheric mass loss \citep[cf.][]{chachan-stevenson2018}, and an additional source of heat from gravitational settling as the core gradually degasses when it cools. 
Depending on its distribution, hydrogen may also produce Ledoux-stable compositional gradients that further insulate the core's primordial heat \citep[cf.][]{scheibe+2021}. 
\\

Understanding thermal evolution in detail is impossible without a better understanding of the coexistence of hydrogen and silicates under the relevant thermodynamic conditions. 
Nevertheless, the timescale for cooling can be estimated to first order from Fig.~\ref{fig:tcools}. 
The cooling of a hydrogen-polluted core will proceed more slowly than suggested by Fig.~\ref{fig:tcools} for two reasons:
first, because as the core degasses the RCB will become deeper, the luminosity will drop further, therefore cooling will become even less efficient. 
Second, the phase separation and settling of higher-density material at depth produces additional heat from gravitational energy. 
A detailed model of this generalized evolution will be the subject of a future work. 
However, we must emphasize that such an endeavor is currently premature. 
A detailed thermal evolution model done properly relies on a better understanding of the physical properties of hydrogen--silicate mixtures under the relevant conditions. 
The stable layer cannot be modeled properly without accurate thermal transport properties of the silicate/hydrogen mixture. 
The behavior of the core cannot be properly modeled without accurate supercritical compressibility data (for example, whether the adiabat is steeper or less steep than the critical curve). 
Thermal evolution cannot be properly modeled without accurate data on the coexistence curve for hydrogen/silicate mixtures. 
It is for these reasons that in this work we restrict ourselves to qualitative commentaries and order-of-magnitude estimations, rather than employing deceptively detailed dynamical models that rely on unsubstantiated assumptions. 
Additional ab intio quantum simulations of these materials under the relevant conditions are of great interest for understanding super-Earths. 
\\

\subsection{Formation and mass loss}
\label{formation}
Our analysis so far has focused exclusively on the planet after formation, and assumes conservation of mass. 
We must include a brief discussion of the dependence of these results on formation conditions. 
The consequential elements to consider are the initial temperature of the core, the initial placement of the RCB, core cooling at early times, atmospheric loss at early times, the distribution of silicate vapor in the envelope, and possible pollution of hydrogen in the core. 
We will discuss each of these points individually. \\

The upper bound for the initial temperature of the core is the accretion energy. 
The gravitational binding energy is $\frac{3 G M_c^2}{5 R_c^2}$. 
Under our assumptions for the relationship between mass and radius for silicate cores, this corresponds to a temperature of approximately 
\begin{equation}
T_{accretion} \sim 2.9 \times 10^4 \left( \frac{M_c}{M_\oplus} \right)^{7/4} \text{K},
\end{equation}
well above the critical temperature for silicate super-Earths. 
However, such extreme temperatures imply extremely efficient cooling according to the Stefan-Boltzmann law. 
Therefore, a detailed formation model is needed in order to determine the actual core temperature at early times. 
Prior modeling of super-Earth planets shows the core temperature can be maintained at high $\sim 10^4$K temperatures at early times after the emplacement of the hydrogen envelope \citep{bodenheimer+2018, brouwers-ormel2020}. 
We point to these results from detailed formation studies to justify our assumption that the core can exist at or above the critical temperature of $\sim 6.6 \times 10^3$K \citep{xiao-stixrude} at early times (see also Fig.~\ref{fig:examples}). 
However, formation models suggest initial core temperatures actually in substantial excess of the critical endpoint temperature, indicating that some initial cooling must occur before before reaching the equilibrium state described in Sect.~\ref{sec:structure}.\\

Indeed, if the core is initially hotter than the critical temperature, or if hydrogen accretions alongside silicates for some portion of the formation process, then this may lead to a situation in which the stable layer truncates according to our arguments in Sects.~\ref{sec:latent} and \ref{sec:coexistence}, and possesses an extended layer of hydrogen-enriched but silicate-dominated material. 
Such a scenario is not precluded by our previous analysis, and the basic arguments still apply.  
However, one must be careful about terminology. 
For example, the total mass of gas in the planet may then exceed what we have been calling the envelope mass. 
Furthermore, evolution would not proceed with the thermal structure of the envelope frozen; rather, the envelope would increase in mass as a function of time. 
In this case, the cooling timescales from Fig.~\ref{fig:tcools} should be thought of as a lower bound, since this scenario includes additional heat sources from further gravitational settling, and longer cooling timescales as the envelope mass increases with time. \\

The RCB at early time will be determined by the disk conditions during formation. 
Formation models find this value to initialize near roughly 10 bars \citep{bodenheimer+2018, brouwers-ormel2020}, varying somewhat depending on specific formation conditions. 
All equilibrium atmospheric profiles must have an RCB at or beneath its formation value, as is the case for higher-mass envelopes. \\

We further did not consider atmospheric loss at early times, a highly complex and poorly understood phenomenon that is likely a diagnostic phase of planetary formation and early evolution. 
This phenomenon has been investigated in detail in prior works, both atmospheric loss by photoevaporation \citep{owen-wu2017} or by core luminosity \citep{gupta-schlichting2019}. 
These dynamics are beyond the scope of this analysis. 
We are not in this work performing detailed evolution models, but rather considering a planet in equilibrium in a snapshot in time. 
We do comment, however, that the core powered mass loss mechanism may be affected when considering convective inhibition, as heat flow from the core may be substantially reduced. 
We also note that hydrogen sequestered in a supercritical core can be a substantial source of additional hydrogen to buffer against envelope mass loss. \\

Our model assumes the envelope is saturated in silicate vapor, and that the core is pure silicates. 
The assumption of a saturated envelope is supported by formation models that explicitly consider silicate vapor \citep{brouwers-ormel2020, bodenheimer+2018}. 
The mechanism for this is the simultaneous accretion of hydrogen and silicates, where at early times accretion is dominated by silicates with some gas pollution, and at later times dominated by gas accretion with silicate pollution ranging from pebbles to planetesimals vaporizing in the atmosphere before cooling and raining out, ensuring a saturated envelope. 
These formation models also find a so-called outer core polluted with hydrogen as a consequence of the stage of formation when some hydrogen is accreted alongside the continued accretion of silicates. 
At later evolutionary phases, this outer core may condense and rain out magma, effectively growing the inner, embryonic core \citep{brouwers-ormel2020}. 
According to our model, we expect a polluted outer core to gradually undergo phase separation as described in Sect.~\ref{core-pollution}. 

\section{Discussion}
\label{sec:discussion}
We have presented a new model for the interior and evolution of super-Earth class planets, likely to be the most common planets in the universe \citep{silburt+2015}. 
Here we have another application of convective inhibition that has been in the literature for decades \citep{guillot1995} and has generated renewed interest in recent years \citep{leconte+2017, friedson-gonzales2017, markham-stevenson2021}. 
In this work we extend the arguments to apply them in the limit where the condensing species rather than the dry gas dominates in abundance. 
We find an extreme case of convective inhibition, wherein a hydrogen atmosphere with a layer of static stability can effectively insulate a core at very high temperatures ($\sim 10^4$K) for geologic time, or potentially longer than the age of the universe. \\

This model should be thought of as exploratory, and further work is necessary. 
To begin with, more work needs to be done to demonstrate the general mechanism of convective inhibition in either a laboratory setting or a high-quality physics-based simulation as an emergent phenomenon. 
Details about our assumptions, including the assumption of saturation and thermodynamic equilibrium, must be assessed in a dynamical context. 
Furthermore, this model must be coupled with a realistic formation scenario, as the subsequent evolution may be sensitive to the planet's initial conditions. 
The stable layer, which for low heat flows may not be extremely thin, must also be considered explicitly along with relevant thermal transport properties. 
Furthermore a serious investigation must explicitly account for the possibility of hydrogen pollution in the core, likely to be substantial based on both formation and thermodynamics arguments. 
Finally, we recapitulate that all of this analysis must rely on additional data on the critical coexistence of hydrogen and silicates, for example from ab initio quantum simulations. 
\\

Our model is of interest for the following reasons. 
First, we predict the luminosity of these planets will be low even at early times, plausibly between the luminosity of Earth and Neptune or less, and should be initially rather insensitive to the planetary system's age for hundreds of millions to billions of years. 
This can potentially be tested via the direct imaging of father out exoplanets in the mass range of interest \citep[e.g.,][]{linder+2019, morley+2017}. 
Second, a supercritical core with a very low internal heat flow may have consequences for the dynamo and expected magnetic field of the planet. 
Third, the temperatures within the core may be so extreme that the contribution to the planet's density due to heat cannot be neglected, and the required quantity of hydrogen to match observed exoplanet radii may be smaller than usually thought \citep[cf.][]{bodenheimer+2018}. 
Along the same lines, sequestered hydrogen may further contribute to core inflation. 
Fourth, the bimodal distribution of exoplanet radii is a highly active area of research. 
This model may be of interest to researchers seeking to explain that observation, because current atmospheric escape models rely on interior models that neglect convective inhibition and the possibility that enormous reservoirs of hydrogen are permitted to dissolve into a supercritical core. 
Finally, this model for super-Earth interiors further demonstrates the importance of considering exoplanetary systems holistically using intuition arising from fundamental physics rather than analogs to more familiar worlds.  
Indeed, it is yet another reminder that our own planet may be atypical. \\

\section*{Acknowledgments}
This work has been primarily funded by NASA FINESST grant number 80NSSC19K1520, as well as funding from the CNES postdoctoral program. 
We would also like to thank Yayaati Chachan for many insightful conversations, and the reviewer for highly constructive feedback. 

\bibliography{library}{}
\bibliographystyle{apalike}

\end{document}